# Trusted Network Selection using SAW and TOPSIS Algorithms for Heterogeneous Wireless Networks

K.Savitha

Research scholar, Periyar University

Salem, Tamilnadu, India

C.Chandrasekar

Associate Professor, Periyar University

Salem, Tamilnadu, India

## ABSTRACT

Seamless continuity is the main goal in fourth generation Wireless networks (FGWNs), to achieve this " HANDOVER" technique is used, when a mobile terminal(MT) is in overlapping area for service continuity, Handover mechanism are mainly used. In Heterogeneous wireless networks main challenge is continual connection among the different networks like WiFi, WiMax, WLAN, WPAN etc. In this paper, Vertical handover decision schemes are compared and Multi Attribute Decision Making (MADM) is used to choose the best network from the available Visitor networks (VTs) for the continuous connection by the mobile terminal. In our work we mainly concentrated to the handover decision phase and to reduce the processing delay in the period of handover. MADM algorithms SAW and TOPSIS where compared to reduce the processing delay by using NS2 to evaluate the parameters for processing delay.

## 1. INTRODUCTION

In fourth generation wireless networks service continuity is a main goal ie., when a MT or mobile node (MN) moving in an overlapping area, continuous service must be need so the technique "HANDOVER" is done. The handover technique [1] is mainly used to redirect the mobile user's service network from current network to a new network or one base station (BS) to another BS or one access point (AP) to another AP with same technology or among different technologies to reduce the processing delay in the overlapping area.

Handover network type (12) has the two types, horizontal handover and vertical handover. The homogenous wireless network performs horizontal handover, if there are two BSs using the same access technology, in current system called horizontal handover. This type of mechanism use signal strength measurements for surrounding BSs to trigger and to perform the handover decision.

In heterogeneous wireless networks environment, always best connected (ABC) [2] which requires dynamic selection of the best network and access technologies when multiple options are available simultaneously.

The mobile station (MS) or BS will be equipped with multiple network interfaces to reach different wireless network. Emerging mix of overlapping heterogeneous wireless networks deployed, vertical handover is used among the networks using different access technologies.

Handover technique has the four phases: Handover Initiation, System discovery, Handover decision, Handoff execution.

- Handoff Initiation phase : The handover process was modified by some criteria value like signal strength, link quality etc.,

- System discovery phase: It is used to decide which mobile user discovers its neighbour network and exchanges information about Quality of Service (QOS) offered by these networks.

- Handover Decision phase: This phase compares the neighbour network QOS and the mobile users QOS with this QOS decision maker makes the decision to which network the mobile user has to direct the connection.

- Handoff Execution phase: This phase is responsible for establishing the connection and release the connections and as well as the invocation of security service.

The scope of our work is mainly in handover decision phase, as mentioned in the decision phase; decision makers must choose the best network from available networks. In this paper, the decision makers are Simple additive weighting (SAW) and Technique for order preference by similarity to ideal solution to take the decision and to select the best target visitor network (TVN) from several visitors' networks.

In this paper, two vertical handover decision schemes (VHDS) , Distributed handover decision scheme (DVHD) and Trusted Distributed vertical handover decision schemes (T-DVHD)are used. DVHD is advanced than the centralised vertical handover decision scheme and T-DVHD is the extended work of DVHD. Here we compare the distributed and trusted vertical handover decision schemes as distributed decision tasks among networks to decrease the processing delay caused by exchanging information messages between mobile terminal and neighbour networks. To distribute the decision task, vertical handover decision is formulated as MADM problem.

In our work, the proposed decision making method use TOPSIS in a distributed manner and compare with SAW method. The bandwidth, delay, jitter and cost are the parameters took by the MT as the decision parameters for handover.

## 2. RELATED WORK

At present many of the handoff decision algorithms are proposed in the literature. In (4) a comparison done among SAW, Technique for Order Preference by Similarity to Ideal Solution(TOPSIS), Grey Relational Analysis (GRA) and Multiplicative Exponent Weighting (MEW) for vertical handoff decision. In (3) author discuss that the vertical handoff decision algorithm for heterogeneous wireless network, here the problem is formulated as Markov decision process. In (5) the vertical handoff decision is formulated as fuzzy multiple attribute decision making (MADM).

In (6) a vertical handoff decision scheme DVHD uses the MADM method to avoid the processing delay. In (7) their goal is to reduce the overload and the processing delay in the mobile terminal so they proposed novel vertical handoff decision scheme to avoid the processing delay and power consumption. In (8) the paper is mainly used to decrease the processing delay and to make a trust handoff decision in a heterogeneous wireless environment using T-DVHD.

In (9) a novel distributed vertical handoff decision scheme using the SAW method with a distributed manner to avoid the drawbacks. In (10) they proposed using the emerging IEEE 802.21 standard defines Media Independent Handover (MIH) functions as transport service in order to offer a vertical handoff decision with a minimum of processing delay. In (11) the paper provides the four steps integrated strategy for MADM based network selection to solve the problem. All of these proposals works mainly focused on the handoff decision and calculate the handoff decision criteria on the mobile terminal side and the discussed scheme are used to reduce the processing delay by the calculation process using MADM in a distributed manner.

In (13) a comparative analysis of MADM methods including SAW, MEW, TOPSIS, ELECTRE, VIKOR, GRA, and WMC is illustrated with a numerical simulation, showing their performance for different applications such as: voice and data connections, in a 4G wireless system.

In (14) , analyzes the advanced tools as well as proven concepts can be used to solve such a problem and thus answering ABC requirement classified the strategies into five main categories: function-based, user-centric, multiple attribute decision, Fuzzy Logic and Neural Networks based, and context-aware strategies. Also compare each one with the others in order to introduce our vertical handover decision approach.

In (16) we compared the three schemes Centralized Vertical handoff decision (C-VHD), Distributed Vertical handoff decision (D-VHD) and Trusted - Distributed Vertical handoff decision (TDVHD). These Schemes provides seamless vertical handoff. The simulation result shows a comparison between three scheme performance in terms of handoff processing delay, end-end delay and throughput.

## 3. TRUSTED VERTICAL HANDOVER DECISION SCHEME

Centralized vertical handover decision (C-VHD), Distributed vertical handover decision (D-VHD), Trusted Distributed vertical handover decision (T-DVHD) are the schemes used to reduce the processing delay between the mobile node and neighbour network while exchanging the information during the handover. In this paper, D-VHD and T-DVHD schemes are compared. MADM have several methods in literature [16]. TOPSIS is used in distributed manner for network selection.

### 3.1 Centralized Vertical Handover Decision Schemes

In C-VHD, a Mobile Node (MN) exchanging the information message to the Neighbour networks mean processing delay was increased by distributing in centralized manner. When processing delay had increased overall handover delay increases. This is one of main disadvantage in C-DHD, so Distributed Vertical handover decision (D-VHD) schemes was proposed in [7][8].

### 3.2 Trusted Distributed Vertical Handover Decision Schemes

D-VHD is used to decrease the processing delay than the C-VHD schemes. This scheme handles the handover calculation to the Target visitor networks (TVNs). TVN is the network to which the mobile node may connect after the handover process was finished. In our work D-VHD takes into account: jitter, cost, bandwidth, delay as evaluation metrics to select a suitable VN which applied in MADM method.

#### 3.2.1 Network Selection Function (NSF):

The network selection decision process has denoted as MADM problem, NSF have used to evaluate from set of network using multiple criteria. The above mentioned parameters are used to calculate NSF. These parameters measure the Network Quality Value (NQV) of each TVN. The highest NQV value of TVN will be selected as Visited Network (VN) by the mobile node. The generic NSF is defined by using SAW "Eq. (3.1) and TOPSIS "Eq. (3.2)"

$$NQV_i = \sum_{i=1, j=1}^{N, n_p^+} W_j * P_{ij} \qquad (3.1)$$

Where, $NQV_i$ represents the quality of $i^{th}$ TVN. $W_j$ is the weight of the $P_{ij}$, $P_{ij}$ represents the $j^{th}$ parameter of the $i^{th}$ TVN. N is the number of TVNS. While $n_p^+$ is the number of parameters.

$$NQV_i = C_i^* = S_i^- / (S_i^* + S_i^-) \qquad (3.2)$$

Where, $NQV_i$ represents the quality of $i^{th}$ TVN. $C_i^*$ is the closeness to the ideal solution.

Based on the user service profile, handover decision parameters have assigns different "Weights" to determine the level of importance of each parameter. In equation (2), the sum of these weights must be equal to one.

$$\sum_{j=1}^{n_p} W_j = 1 \qquad (3.3)$$

The handover decision metrics calculation is performed on the VNs, each VN applies the MADM methods using "Eq. (3.1,3.2)" on the required ($J_{req.}$, $D_{req.}$, $C_{req.}$, $B_{req}$) and offered ($J_{off}$, $D_{off}$, $C_{off}$, $B_{req}$) parameters

### 3.2.2    Distributed Decision scheme:

The D-VHD is explained in the Figure 1. Therefore, the DVHD scheme consists on the following steps:

- The mobile node initiates the handoff process, caused by the degradation of the offered quality or the availability of TVNs offering better quality then the quality offered by the network to which the mobile node is connected. Then it sends a handoff request message to all available TVNs, this message includes the mobile node identity and the user profile reference.

- Each TVN computes its NQV, by retrieving the appropriate User-Profile table, then it creates the decision matrix and the weight on the required ($J_{req.}$, $D_{req.}$, $C_{req.}$, $B_{req}$) and offered ($J_{off}$, $D_{off}$, $C_{off}$, $B_{req}$) parameters .Then it sends its NQV to the mobile node.

- Finally, the mobile node puts all received NQVs in a list, then it picks up the highest NQV and considers that the corresponding TVN is the VN, to which it redirects all connections.

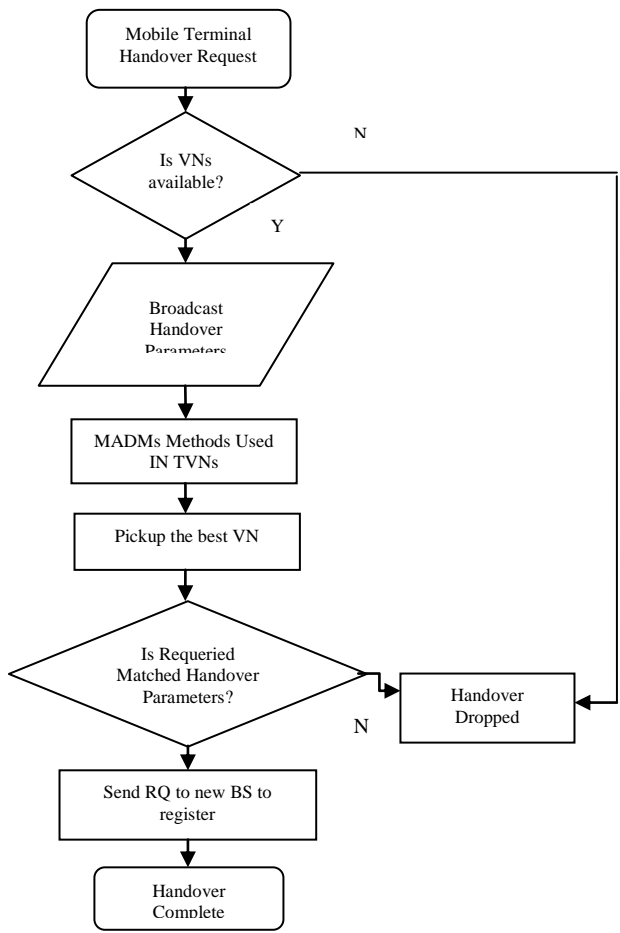

**Fig. 1 D-VHD Scheme**

### 3.2.3    Trusted Distributed Decision schemes

Trusted handover decision and to avoid the unnecessary handover events are the important factors while exchanging the trusted information between networks and mobile node. The extension work of the DVHD scheme is T-DVHD scheme. The scheme is mainly introduced [10] for decreasing the processing delay than DVHD scheme.

- The T-DVHD schemes followed by the DVHD Network selection function and Distribute Decision schemes, before sending request to connect a new base station trusted process is started

#### 3.2.3.1   Level Of Trust (LOT) test function

LOT function is tested to execute the handover. LOT function is calculated by the following steps

```
If LoTi >= threshold
        Connect to the TVNi
        start Trust-test function
else if LoTi < threshold {
        if (suitable-TVN available)
                i = i + 1
                test another network
        else if (no suitable-TVN)
Handover blocked
```

after handover is executed by the mobile terminal with the proper TVN. Trusted Test Function is started, once the mobile terminal connects to the TVN trusted test function is calculated by the following steps to finish the T-DVHD schemes.

if Qoff < Qreq

$$LOT_i = LOT - delta \; ;$$

else

$$LOT_i = LOT_i + delta^+ \; ;$$

## 4.    DECISION MAKERS FOR VERTICAL HANDOVER DECISION SCHEMES

Multiple attribute decision making (MADM) refers to making preference decisions (e.g., evaluation, prioritization, and selection) over the available alternatives that are characterized by multiple, usually conflicting, attributes. The structure of the alternative performance matrix " Table 1",where xij is the rating of alternative i with respect to criterion j and wj is the weight of criterion j. Since each criterion has a different meaning, it cannot be assumed that they all have equal weights, and as a result, finding the appropriate weight for each criterion is one the main points in MADM. Various methods for finding weights can be found in the literature and most of them can be categorized into two groups: subjective and objective weights. Subjective weights are determined only according to the preference decision makers. The objective methods determine weights by solving mathematical models without any consideration of the decision maker's preferences.

Table 1: Matrix format a MADM problem

|     | $C1(w1)$ | $C2(w2)$ | $C3(w3)$ | $C4(w4)$ |
|-----|------|------|------|------|
| $A1$ | $x11$ | $x12$ | $x13$ | $x14$ |
| $A2$ | $x21$ | $x22$ | $x23$ | $x24$ |
| ... | ... | ... | ... | ... |
| ... | ... | ... | ... | ... |
| ... | ... | ... | ... | ... |
| $Ai$ | $xi1$ | $xi2$ | $xi3$ | $xi4$ |

In this paper, we have compared two decision makers SAW and TOPSIS for VHDS as distributed manner.

In Figure 2 network ranking module integrates all the information coming from weighting and adjusting modules, and obtains a rank of all the networks.

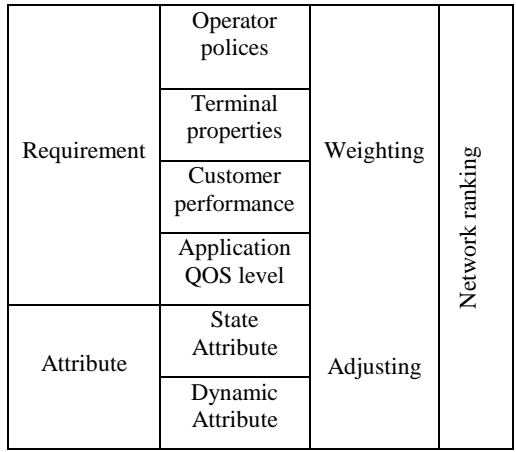

Fig .2 MADM Based Network Selection

MADM algorithms that have been used for network ranking include simple additive weighting (SAW), multiplicative exponential weighting (MEW), technique in order to preference by similarity to ideal solution (TOPSIS), grey relational analysis (GRA), elimination and choice translating reality (ELECTRE) etc., The first four algorithms rank networks based on their coefficients calculated b combining adjusted values of all the criteria.

## 4.1    Simple Additive Weighting (SAW) Method:

Simple Additive Weighting (SAW) which is also referred as weighted linear combination or scoring methods or weighted sum method is a simple and most often used multi attribute decision technique. The method is based on the weighted average. An evaluation score is calculated for each alternative by multiplying the scaled value given to the alternative of that attribute with the weights of relative importance directly assigned by decision maker followed by summing of the products for all criteria.

The application of SAW scoring requires , identification of objectives and alternatives, evaluation of alternatives, determination of sub-objective weights, additive aggregation of weighted partial preference values, sensitive analysis. It uses direct rating on the standardised scales only in purely qualitative attributes. For numerical attributes score are calculated by normalized values to match the standardised scale.

The SAW is a comparable scale for all elements in the decision matrix, the comparable scale obtained by $r_{ij}$ for benefit criteria "Eq. (4.1)" and worst criteria "Eq.(4.2)" .

$$V_{ij} = \frac{x_{ij}}{x_j^{max}} \qquad (4.1)$$

$$V_{ij} = \frac{x_j^{min}}{x_j} \qquad (4.2)$$

The SAW method , underlying additive values function and compute as alternatives score $V_i = V(A_i)$ by adding weighting normalized values $W_j V_{ij} \, \forall \, j = \{1,\ldots\ldots.m\}$ before eventually ranking alternatives

$$V_i = \sum_{j=1}^{m} W_j V_{ij} \qquad (4.3)$$

For V ε $R^{n*m}$ with i = { 1,……,n}, j = {1,……..,m}; $V_{ij}$ , $W_j$ ε (0,1)

## 4.2    Technique for Order Preference By Similarity To Ideal Solution (Topsis)

TOPSIS (15) is a MADM instrument for measuring relative efficiency of alternatives. It determines the preference order on the grounds of the similarity to a positive ideal solution and the worst similarity to a negative solution. The following are the steps of TOPSIS.

Construct the normalized decision matrix. Each element $rij$ of the Euclidean normalized decision matrix $R$ can be calculated as follows:

$$r_{ij} = \frac{x_{ij}}{\left(\sum x_{ij}^2\right)} \qquad \text{for i=1,…,m;} \qquad \text{j=1,……,n}$$

$$(4.4)$$

Next the weighted normalized decision matrix is constructed by

$$V_{ij} = w_j r_{ij} \qquad (4.5)$$

Then positive ideal and negative ideal solutions are determined by

Positive Ideal solution.

$$A^* = \{v_1^*, \ldots\ldots\ldots, v_n^*\} \quad \text{where} \qquad (4.6)$$

$$V_j^* = \{max_i(V_{ij}) \, if \, j \in \text{J} \, ; min_i(V_{ij}) \, if \, j \in \text{J}^{'}$$

Negative ideal solution.

$$A^{'} = \{v_1^*, \ldots\ldots\ldots, v_n^*\}, \quad \text{where} \qquad (4.7)$$

$$V_j^* = \{min_i(V_{ij}) \, if \, j \in \text{J} \, ; max(V_{ij}) \, if \, j \in \text{J}^{'}$$

The distance between each alternative and the positive ideal solution is:

$$S_i^* = [\sum_j (v_j^* - v_{ij})^2]^{1/2} \text{ i = 1, …, m} \qquad (4.8)$$

The distance between each alternative and the negative ideal solution is:

$$S_i^{'} = [\sum_j (v_j^{'} - v_{ij})^2]^{1/2} \quad i = 1, \dots, m \qquad (4.9)$$

Finally relative closeness to the ideal solution $C_i^{*}$ is calculated as

$$C_i^{*} = S_i^{'}/(S_i^{*} + S_i^{'}) , \quad 0 < C_i^{*} < 1 \quad (4.10)$$

## 5. NUMERICAL EXAMPLE

The above section outlines the vertical handover decision schemes and MADM methods, SAW and TOPSIS which is used for the network selection in this paper. For instance, suppose a mobile terminal is currently connected to a WiFi cell and has to make decision among six candidate networks A1, A2, A3, A4, A5, A6, where A3, A4 are WiFi cells and others are WiMax cells. Vertical handover criteria considered here are delay, bandwidth, cost, jitter which denoted as X1, X2, X3,X4 respectively. Decision matrix D is as follows

$$D= \quad \begin{array}{ccccc} & X1 & X2 & X2 & X2 \\ A1 & 0.00062 & 8 & 9 & 0.411 \\ A2 & 0.00063 & 1.5 & 8 & 0.762 \\ A3 & 0.00062 & 15 & 12 & 0.057 \\ A4 & 0.00063 & 7 & 6 & 0.939 \\ A5 & 0.00062 & 11 & 10 & 0.103 \\ A6 & 0.00061 & 1 & 9 & 0.247 \end{array}$$

The users running application was voice. The preference on handover criteria is modelled as weights assigned by the user on the criteria, for voice $W_v$ which shown in the "Eq. (5.1)".

$$W_v = [0.3 \ 0.2 \ 0.2 \ 0.3] \qquad (5.1)$$

MADM methods handle in this paper for decision problems with above data. The following section discussed about the SAW and TOPSIS are applied and the results are compared.

## 5.1 SAW

SAW requires a comparable scale for all elements in the decision matrix, the comparable scale is obtained by using "Eq. (4.1), Eq. (4.2)". In these $x_{ij}$ is the performance score of alternatives $A_i$ with respect to criteria $x_j$. after scaling, the normalized decision matrix is evaluated as $D^{'}$

$$D^{'}= \quad \begin{array}{ccccc} & X1 & X2 & X2 & X2 \\ A1 & 0.984 & 0.533 & 1.5 & 0.438 \\ A2 & 1 & 0.1 & 1.33 & 0.812 \\ A3 & 0.984 & 1 & 2 & 0.061 \\ A4 & 1 & 0.467 & 1 & 1 \\ A5 & 0.984 & 0.733 & 1.67 & 0.119 \\ A6 & 0.968 & 0.667 & 1.5 & 0.263 \end{array}$$

Applying the weight factor from the "Eq. (5.1)", weighted average values for A1, A2, A3, A4, A5 and A6 are calculated for the respected to the voice application $A_v$

$$A_v = \quad 0.833 \quad 0.83 \quad 0.813 \quad 0.893 \quad 0.809 \quad 0.802$$

The best network is A4 which is the network selected to connect the mobile terminal for service continuity with the minimum processing delay.

## 5.2 TOPSIS

Using TOPSIS, the first step is to construct normalized decision matrix

$$r_{ij} = \quad \begin{array}{ccccc} & X1 & X2 & X3 & X4 \\ A1 & 0.413 & 0.372 & 0.4 & 0.314 \\ A2 & 0.412 & 0.069 & 0.36 & 0.583 \\ A3 & 0.413 & 0.678 & 0.53 & 0.436 \\ A4 & 0.42 & 0.326 & 0.27 & 0.718 \\ A5 & 0.413 & 0.512 & 0.44 & 0.079 \\ A6 & 0.407 & 0.047 & 0.4 & 0.189 \end{array}$$

The decision matrix for voice is weighted using the weighting factors from $W_v$ and the weighted normalized matrix $V_{ij}$ is

$$V_{ij}= \quad \begin{array}{ccccc} & X1 & X2 & X3 & X4 \\ A1 & 0.124 & 0.074 & 0.08 & 0.094 \\ A2 & 0.126 & 0.014 & 0.07 & 0.175 \\ A3 & 0.124 & 0.136 & 0.11 & 0.131 \\ A4 & 0.126 & 0.098 & 0.05 & 0.215 \\ A5 & 0.124 & 0.102 & 0.09 & 0.024 \\ A6 & 0.122 & 0.009 & 0.08 & 0.057 \end{array}$$

To determine the positive ideal solution $A^{*}$ and negative ideal solution $A^{-}$

$$A^{*} = 0.126 \quad 0.136 \quad 0.5 \quad 0.215$$

$$A^{-} = 0.122 \quad 0.009 \quad 0.11 \quad 0.024$$

To determine the distance between each alternative and the positive ideal solution and negative ideal solution

$$S_i^{*}= \begin{array}{c} 0.139 \\ 0.129 \\ 0.103 \\ 0.038 \\ 0.198 \\ 0.205 \end{array} \quad S_i^{-}= \begin{array}{c} 0.100 \\ 0.156 \\ 0.166 \\ 0.219 \\ 0.095 \\ 0.045 \end{array}$$

Finally $C_i^{*}$ closeness of the ideal solution shows .From $C_i^{*}$, A4 base station is the best to connect the mobile terminal by TOPSIS decision maker

$$C_i^{*} = \quad \begin{array}{cc} A1 & 0.42 \\ A2 & 0.55 \\ A3 & 0.62 \\ A4 & 0.85 \\ A5 & 0.32 \\ A6 & 0.18 \end{array}$$

## 5.3 Comparison of MADM method

The ranking order using different methods of MADM are summarised in "Table 2". SAW and TOPSIS ranks A4 is the best to handover to the new base station, because in SAW A4 has good scores on jitter, cost, delay and in TOPSIS A4 has good scores on jitter and delay. So the A4 BS have connect the mobile terminal with less processing to get seamless handover in between the MT and BS A4

Table 2: Ranking order comparison

| | | | | | | |
|---|---|---|---|---|---|---|
| *SAW* | A4 | A1 | A2 | A3 | A5 | A6 |
| *TOPSIS* | A4 | A3 | A2 | A1 | A5 | A6 |

# 6. SCENARIO OF THE VERTICAL HANDOVER

In this paper, our scenario was in "Figure 3", it explains that a cell coverage the area by WiMax technology and another cell coverage the area by WiFi and WiMax technology. A mobile terminal is overlapping with VoIP application between the cell coverage now mobile terminal intend to connect the appropriate visited network with the decision process.

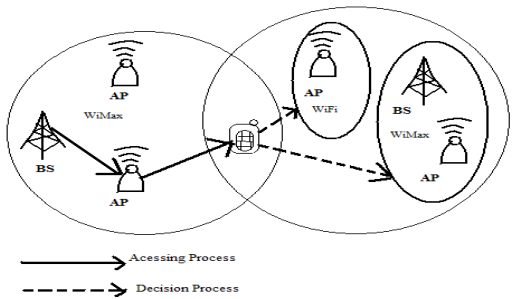

Fig. 3 Scenario of the vertical handover

# 7. SIMULATION

In this section, the comparisons of vertical handover decision scheme are compared and we provide the evaluation parameters used to analyze the performance T-DVHD schemes as well as the output of simulation. In our simulation we consider 7 mobile nodes are moving in an area covered by the heterogeneous wireless networks managed by 6 Base stations . Mobility area covered by BS, supporting two types of technologies: WiMax and WiFi. These BS offer different characteristic in terms of coverage and QOS . VoIP is used as application in this simulation.

## 7.1 Evaluation Parameters

There are different evaluation parameters are used, in order to evaluate our schemes. We have used:

- Processing Delay: It is a process which takes time by the terminal for making the decision towards which network to handover for network to handover

- Throughput: It is measured by the data are sent by the mobile node after a set of matching decision during a defined period.

- End to End Delay: It refers the time taken for a packet to be transmitted across a network from source to destination

- Handover Events: It reflects the number of handover achieved by the mobile terminal

- Packet Delivery Ratio : It defined as the number of received  data packets divided by the number of generated data packets

## 7.2 Simulation Analysis

In Simulation Analysis, "Figure 4" "Figure 5" "Figure 6" shows the processing delay of different Visitor networks like 2VN, 3VN, 4VN. The processing delay time is taken in seconds. The time has taken for completing the whole handover process is analyzed in this Process Delay. In "   "   the comparison of CVHD, DVHD and T-DVHD are shown and from that T-DVHD is analyzed as the best from vertical handover decision schemes. In this paper, evaluation parameters used to analyze the performance of T-DVHD scheme.

The processing delay analyze for different visitor network show that TOPSIS is good decision maker than SAW in less processing delay for handover.

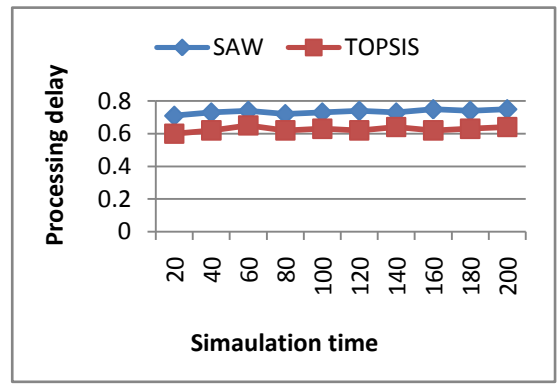

Fig . 4 Handover processing delay between 2VN

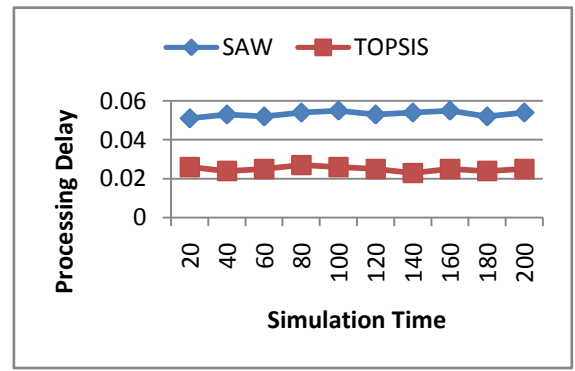

Fig.5 Handover processing delay among 3 VN

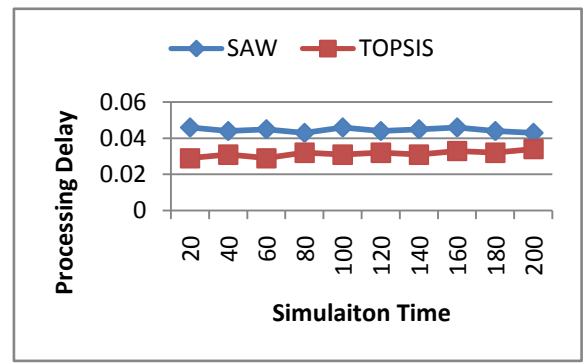

Fig.6 Handover processing delay among 4 VN

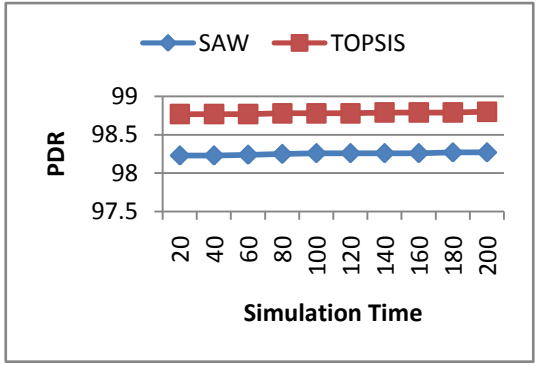

Fig.7 Packet Delivery Ratio between 2VNs

In "Figure 7" "Figure 8" "Figure 9" show the packet delivery ratio for different Visitor networks like 2VNs, 3VNs, 4VNs.

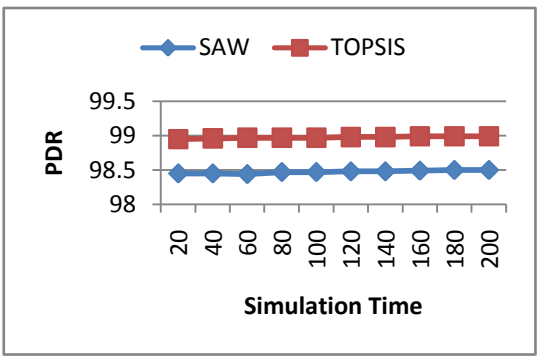

Fig.8 Packet Delivery Ratio between 3VNs

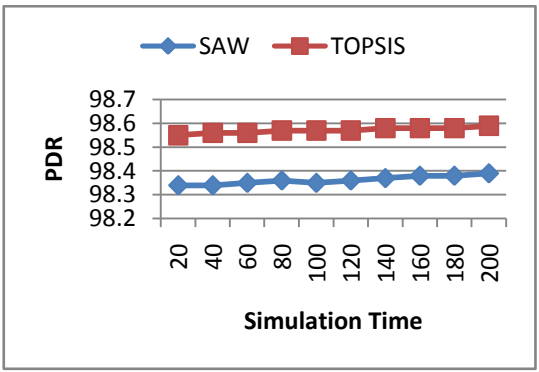

Fig.9 Packet Delivery Ratio among 4VNs

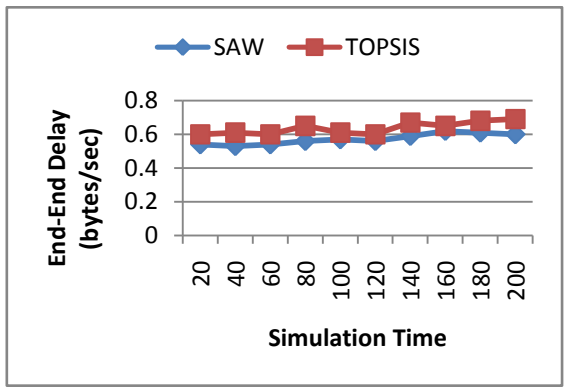

Fig.10 End to End Delay

In "Figure 10", it explained that End to End delay between the node and destination access point with required QOS service. End -End delay is sum of transmission delay, propagation delay and processing delay of number of links. In this SAW is better than TOPSIS

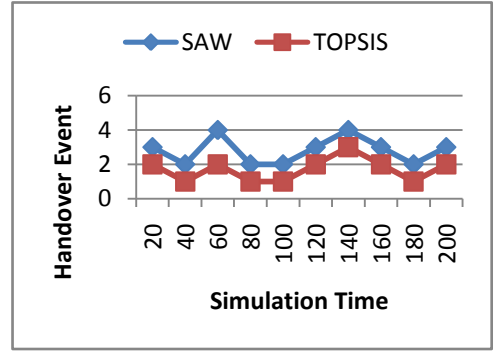

Fig.11 Handover Events

In "Figure 11" multiple handover events are occurred, when the mobile node chooses a TVN that provides falsified quality value (i.e. NQV). In case, another handover event may be performed as the switched VN doesn't provide the appropriate quality, which adds additional delay to the handover process.

Throughput in "Figure 12" shows by the mobile terminal. Throughput is measured in bits per second. It calculated by Total Bytes Sent * 8 divide by Time Last Packet Sent - Time First Packet Sent here time is in seconds. This shows that TOPSIS is a good decision maker than the SAW

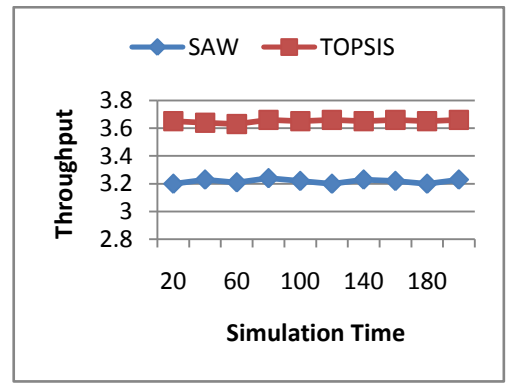

Fig.12 Mobile terminal throughput

# 8. CONCLUSION

In our work, we have compared the schemes of vertical handover decision in the heterogeneous wireless networks. The observation of schemes to reduce the processing delay and a trusted handover decision is done in heterogeneous wireless networks. In this paper we proposed decision maker TOPSIS to select the best network from the visitor network for the Vertical decision schemes. The simulation analyze shows TOPSIS is best decision maker than SAW to select the best network to handover for target visitor network. Our main goal is in the decision phase of the handover phases to take decision to which VN the mobile terminal to connect to decrease the processing delay by different decision algorithms

# 9.     REFERENCES

[1] N.D.Tripathi, J.H.Reed and H.F.VanLandinoham, , "Handoff In Cellular System", *IEEE Personal Communications*, Vol.49, 2000, pp.2276-2285,

[2] E.Gustafsson and Annika Jonsson, , Always Best Connected, *IEEE Wireless Communications,* Vol.10,No. 1, 2003, pp. 49-55,

[3] E.Steven-Navarro, V.W.S. Wong and Yuxia Lin, "A Vertical Handoff Decision Algorithm For Heterogeneous Wireless Networks", *Wireless Communications and Networking Conference, IEEE,* Kowloon 2007, pp. 3199-3204.

[4] E.steven-Navarro and V.W.S.Wong, , "Comparison between vertical handoff decision algorithms for heterogeneous wireless network", *Vehicular Technology Conference, IEEE 63rd* ,Melbourne, Vic. , 2006, pp.947-951.

[5] Wenhui Zhang, " Handover Decision Using Fuzzy MADM In Heterogeneous Networks", *Wireless Communications and Networking Conference, IEEE,* Vol.2, 2004, pp. 653-658.

[6] R.Tawil, G.Pujolle and O.Salazar, , A Vertical "Handoff Decision Schemes In Heterogeneous Wireless Systems", *Vehicular Technology Conference, VTC Spring 2008. IEEE* , Singapore , 2008,pp. 2626-2630.

[7] R.Tawil, J.Demerjain, G.Pujolle And O.Salazar, "Processing-Delay Reduction During The Vertical Handoff Decision In Heterogeneous Wireless System" *International Conference on Computer Systems and Applications, AICCSA IEEE/ACS 2008,* pp.381-385.

[8] R.Tawil, J.Demerjain and G.Pujolle, "A Trusted Handoff Decision Scheme For The Next Generation Wireless Networks", *IJCSNS*, Vol.8, 2008, pp.174-182.

[9] R.Tawil, G.Pujolle and O.Salazar, "Vertical Handoff Decision Schemes For The Next Generation Wireless Networks", *Wireless Communications and Networking Conference,* 2008, pp. 2789-2792.

[10] R.Tawil, G.Pujolle and J.Demerjain, "Distributed Handoff Decision Scheme Using MIH Function For The Fourth Generation Wireless Networks", *3rd International Conference on Information and Communication Technologies: From Theory to Applications,* 2008, pp. 1-6.

[11] Lusheng Wang and David Binet, "MADM- Based Network Selection In Heterogeneous Wireless Networks: A Simulation Study", *1st International Conference on Wireless Communication, Vehicular Technology, Information Theory and Aerospace & Electronic Systems Technology,* 2009, pp. 559-564.

[12] N.Nasser, A.Hasswa and H.Hassanein, "Handoff In Fourth Generation Heterogeneous Networks", *Communications Magazine, IEEE* , Vol.44, 2006, pp. 96-103.

[13] J.D Martinez- Morales, V.P.rico, E.Steven, "Performance comparison between MADM algorithms for vertical handoff in 4G networks", *7th International Conference on Electrical Engineering Computing Science and Automatic Control (CCE),* 2010, pp. 309-314.

[14] M.Kassar, B.Kervella, G.Pujolle, "An overview of vertical handover decision strategies in heterogeneous wireless networks", *Elsevier, Journal of computer communications*, Vol.37, No.10, 2008.

[15] D.L.Olson, "Comparison of weights in TOPSIS models', *J Mathematical and computer modelling*, 2000, pp. 40721-727.

[16] K.Savitha, C.Chandrasekar, "Comparison of Vertical Handoff Decision Scheme in Heterogeneous Wireless Network", *IEEE International Conference on Computational Intelligence And Computing Research,* 2010, pp. 1-6.